\documentstyle[11pt,paspconf,epsf]{article}

\begin{document}

\title{Bulk-flow and $\beta_I$ from the SMAC project}
\author{Russell J. Smith$^1$, Michael J. Hudson$^2$, John R. Lucey$^1$, \\
David J. Schlegel$^3$ and Roger L. Davies$^1$}
\affil{$^1$\,Department of Physics, University of Durham, 
Science Laboratories, South Road, Durham, DH1 3LE, United Kingdom.}
\affil{$^2$\,Department of Physics \& Astronomy, University of Victoria, \\
PO Box 3055, Victoria BC V8W 3PN, Canada.}
\affil{$^3$\,Department of Astrophysical Sciences, Princeton University, \\
Peyton Hall, Princeton NJ, USA.}

\begin{abstract}
The SMAC project is a Fundamental Plane peculiar velocity survey of 56 
clusters of galaxies to a depth of $cz\sim$12000\,km\,s$^{-1}$.
We present here some results from the analysis of the SMAC velocity field, 
focussing on three specific features: the 
best-fitting bulk-flow model for the SMAC data; the agreement between the 
observed velocity field and predictions from the IRAS--PSCz redshift survey;
the role of the Great Attractor and Shapley Concentration in generating the 
local flows. We argue that the local mass distribution, as probed by the PSCz, 
can fully account for the observed cluster velocities.
\end{abstract}

\section{Introduction}

The `Streaming Motions of Abell Clusters' (SMAC) project is a Fundamental Plane (FP) 
survey of $\sim$700 early-type galaxies in 56 local rich clusters. The cluster sample has 
approximately full-sky coverage, and a limiting depth of $\sim12000$\,km\,s$^{-1}$.
Within each cluster, distances are based upon data for 4--56 E/S0 galaxies. 
Data for the SMAC project is drawn from a compilation of literature sources and an extensive
body of new observations, carefully combined into a homogeneous database. Further details of
the sample, the observations, and the data compilation procedures are reported
by Hudson et al. (1999) and in a forthcoming series of papers.

Our (inverse) FP analysis closely follows the methods employed, for a smaller sample of clusters,
by Hudson et al. (1997). The FP yields distances to
a precision of $\sim20\%$ per galaxy, so that cluster distances are determined with errors
of 3--12\%, with a median 8\%. The velocity zero-point is calibrated by requiring that
the sample exhibit no net radial inflow or outflow. The balanced sky-coverage of the
sample ensures that there is little covariance between monopole and dipole components 
of the velocity field. 
The observed velocity field is presented in the upper panel of Figure~1.
Lucey et al. 
(in this volume) 
present
some example comparisons of the SMAC distances/velocities
with those determined from other surveys with clusters in common (eg Lauer \& Postman 1994; 
Dale et al. 1999). Such comparisons generally reveal an 
acceptable level of agreement, in so much as the scatter 
is compatible with the quoted errors.
However, the distance errors are such that cluster-by-cluster comparisons 
are rather crude tests for systematic effects in distance estimates.

\begin{figure}\label{galplot}
\vskip -15mm
\plotone{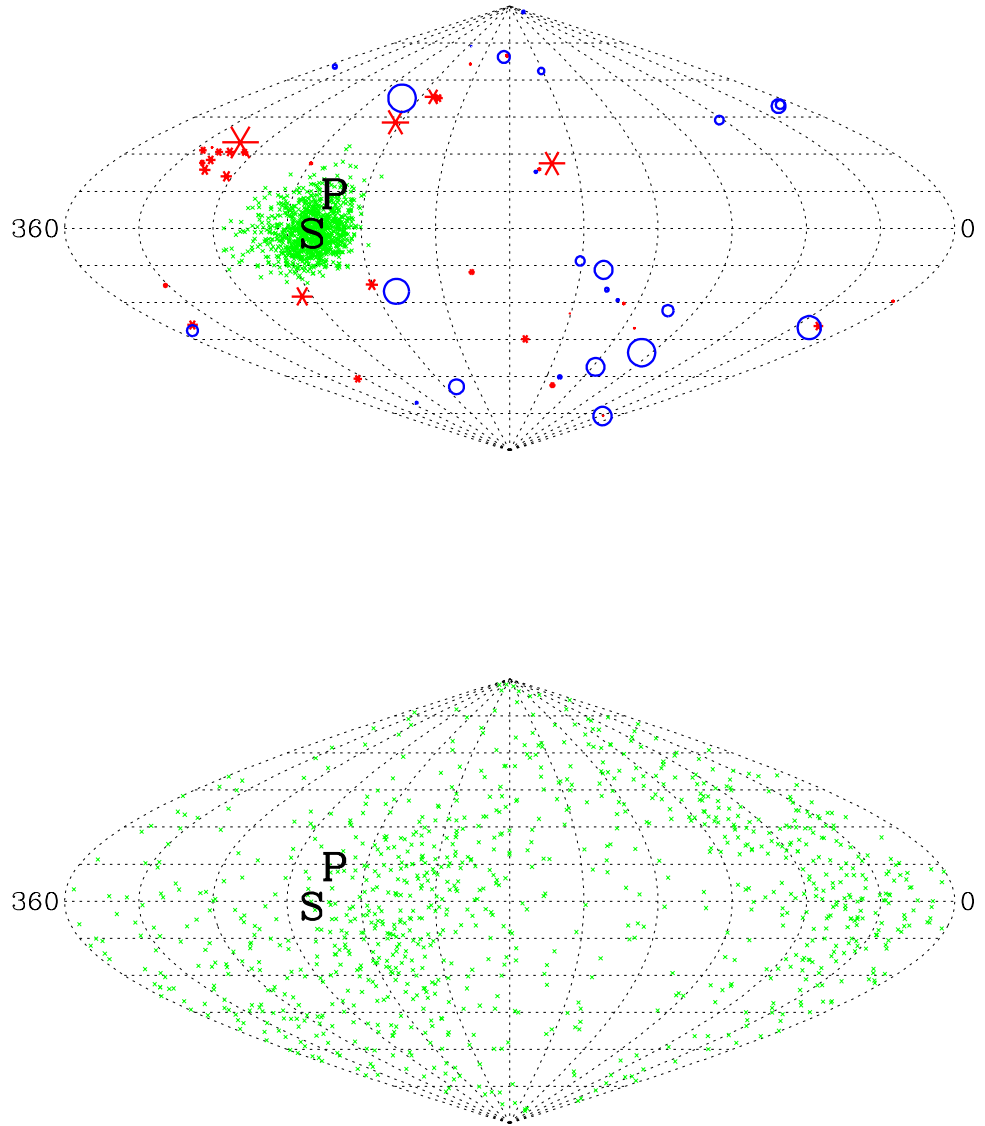}
\vskip -10mm
\caption{The upper panel shows the sky-projection of the SMAC peculiar velocity field
(CMB frame), in galactic co-ordinates. Clusters with positive (negative) peculiar motions are
indicated by asterisks (circles), with symbol sizes are assigned according 
to the magnitude of the cluster peculiar velocity.  
The {\bf S} indicates the direction of the observed bulk-flow 
(ie the vector ${\bf V}_B^{obs}$) while {\bf P} shows the PSCz-prediction for the 
bulk-flow (${\bf V}_B^{pred}$). The cloud of small crosses show the directional error 
determined from Monte-Carlo simulations.
In the lower panel, we show the bulk-flow direction computed from 1000 Monte-Carlo simulations in 
which the true cluster velocities are all zero. The preference for directions 
$b\sim0^\circ, l\sim50^\circ,230^\circ$ results from the relatively poor sampling of 
this axis by the cluster distribution. 
}
\end{figure}

\section{The SMAC bulk-flow}

The best-fitting bulk-flow model is the vector ${\bf V}_B$ which minimizes 
\begin{equation}
\chi^2=\sum_i\frac{(v_i - {\bf V}_B.\hat{\bf r}_i)^2}{\sigma_i^2+\sigma_{\rm th}^2}\, ,\label{chibulk}
\end{equation}
where $v_i$ is the observed radial peculiar velocity for cluster $i$, whose
direction vector is $\hat{\bf{r}}_i$. The weights are assigned according to the 
observational peculiar velocity error, $\sigma_i$, and a `thermal' velocity noise
$\sigma_{\rm th}$, here fixed at 250\,km\,s$^{-1}$. In what follows, we shall refer to
${\bf V}_B$ as `the bulk-flow', but this terminology should be understood as a 
shorthand for `the best-fitting bulk-flow model'. As demonstrated by Feldman \& Watkins (1994)
and further illustrated by Hudson et al. (this volume), the incomplete cancellation of 
small- and intermediate-scale flows causes bulk-flow measurements to depend 
upon the sample geometry, characterised by the survey window function. 
This effect is especially important when the spatial sampling is sparse, as for cluster 
surveys. 

For the SMAC sample, we determine the following CMB-frame bulk-flow solution: 
\begin{equation}\label{obsbulk}
{\bf V}_B^{obs} = [ -345\pm85 , +37\pm101 , -538\pm158 ]\,\, {\rm km\,s}^{-1}
\end{equation}
in supergalactic cartesian co-ordinates (used throughout this paper for flow vector components).
After correcting for `error-biasing', this vector has magnitude $640\pm200$\,kms$^{-1}$
and is directed towards $(l,b) = (260^\circ, -2^\circ)$. Note that the error ellipsoid
is anisotropic, as demonstrated by Monte-Carlo simulations in the lower panel of 
Figure~1. The bulk-flow errors include a 
contribution from `system-matching' errors which introduce covariance between cluster distance
errors (Smith et al. 1997; Hudson et al. 1997). 

The large amplitude of the observed flow is initially surprising. 
We have tested for a wide range 
of possible systematic effects which might affect this result: no single cluster, or
supercluster region, is responsible for the flow; correcting FP distances for stellar population
differences, based on the clusters' offsets from the Mg$-\sigma$ relation, 
does not significantly affect the bulk-flow; our choice of Schlegel et al. (1998) extinction
maps over Burstein \& Hieles (1982) makes little difference to the result; 
a `Method II' analysis, which fits simultaneously the FP parameters and flow
model and is insensitive to Malmquist Bias, yields indistinguishable results. 

At face value, the SMAC result appears inconsistent with the bulk-flow solution of
Lauer \& Postman (1994), whose apex lies $\sim90^\circ$ from the SMAC flow direction, 
as well as with the non-detection of bulk-motion in the Tully--Fisher survey of Dale 
et al. (1999, and these proceedings). 
In fact, the inconsistencies between surveys are marginal when one accounts for differences
in the survey window functions (again see Hudson in this volume). For the same reason, the
apparently good agreement between the SMAC and Willick (1999) solutions should not be
over-interpreted: 
the surveys have very different window functions, and we should not {\it expect} the 
derived bulk-flows to agree so closely!

\section{Comparison to IRAS--PSCz}

How to interpret the SMAC bulk-flow result depends upon whether the observed
velocities of clusters (individually, or combined via the best-fitting bulk-flow 
model) can be understood as the expected response to the surrounding density field.
Here, we compare the SMAC data to density-velocity reconstructions from the most
extensive all-sky redshift survey of IRAS galaxies, the PSCz
(see contributions of Saunders and Branchini in this volume). 
The analysis presented here is somewhat preliminary, and various systematic effects
remain to be investigated.

First, let us consider the predictions for the bulk-flow. We compute predicted velocities
for each of the SMAC clusters (assuming $\beta_I=\Omega_0^{0.6}/b_I=1$ for now), and substitute
these predictions in place of the $v_i$ in Equation~\ref{chibulk}. We again minimize $\chi^2$ with respect to 
${\bf V}_B$, using the same weights computed from the SMAC measurement errors and the thermal velocity 
noise. The best-fitting flow vector is what PSCz predicts {\it for the SMAC bulk-flow}, ie. given
the SMAC errors and the SMAC sample geometry. 
Specifically, we fully account for any over-representation of clusters in (for example) 
the Great Attractor direction. 
The {\it expected} bulk-flow solution for $\beta_I=1$ is then: 
\begin{equation}\label{predbulk}
{\bf V}_B^{pred} = [ -209, +195, -475 ]\,\, {\rm km\,s}^{-1}\, .
\end{equation}
A cursory comparison between Solutions~\ref{obsbulk} and \ref{predbulk} reveals that the
PSCz velocity field for $\beta_I=1$, given the SMAC sampling geometry and measurement errors, 
predicts a best-fitting bulk-flow similar to that observed, (including the large negative 
SGZ component). Solution~\ref{predbulk} has magnitude 490$\pm$140\,km\,s$^{-1}$  
(error-bias corrected) and direction $(l,b)=(253^\circ,14^\circ)$, which is only $\sim$17$^\circ$ 
from the direction of the observed flow apex (see Figure~1). 
From Monte-Carlo simulations (lower panel of Figure~1), we find that such a good {\it directional} 
agreement arises by chance in $<3\%$ of realisations. 

Alternatively, we can compare directly the predicted and observed velocity fields on a 
cluster-by-cluster basis, as in Figure~2. The fit
yields a measurement of $\beta_I=0.95\pm0.19$, with no individual cluster influencing the 
result by more than 10\%. We can introduce greater freedom into the model velocity field
by fitting simultaneously for $\beta_I$ and a `residual' bulk-flow vector. This residual 
bulk-flow will absorb any
dipole signature {\it not} accounted for by the PSCz galaxy distribution (either a real
streaming generated at very large depths, or a spurious flow resulting from systematic
errors). The results of the fit are an unchanged value for $\beta_I=0.94\pm0.25$ and a residual
flow vector:
\begin{equation}\label{residbulk}
{\bf V}_B^{resid} = [ -169\pm103 , -117\pm110 , -170\pm190 ]\,\, {\rm km\,s}^{-1}\, .
\end{equation}
The errors are larger than previously since we are now fitting for more parameters; however, 
the correlation of observed with predicted peculiar velocities remains significant at
the $>3\sigma$ level.
That the residual bulk-flow is not significant is a strong indication that the observed 
SMAC flow is indeed compatible with the expected response to local density fluctuations; 
and suggests that the observed flow does not result primarily from systematic distance 
errors. The favoured value for $\beta_I$ is higher than many estimates, but is consistent 
with other determinations based on PSCz, which yield $\beta_I=0.6-0.7$ (see Branchini, 
this volume). Note that there is some covariance between  $\beta_I$ and ${\bf V}_B^{resid}$, 
such that fixing  $\beta_I=0.7$ results in a solution with larger residual flow, albeit
significant at only the 1.5$\sigma$ level.

\begin{figure}
\vskip -8mm
\plotone{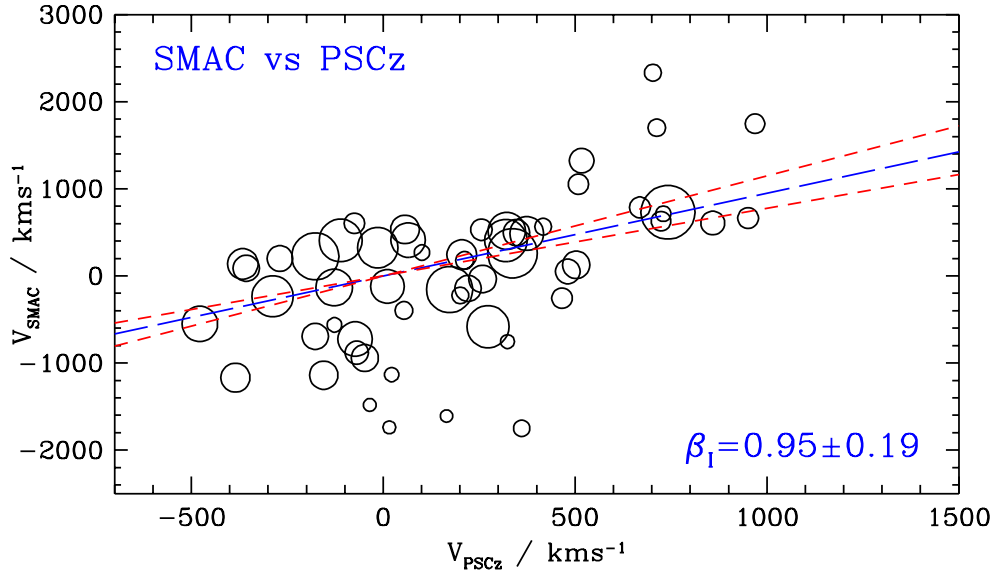}
\vskip -95mm
\caption{Cluster-by-cluster comparison of SMAC-measured peculiar velocities to the radial
peculiar velocities predicted from IRAS--PSCz (for $\beta_I$=1). 
Symbol sizes are proportional to the weight carried in the fit for $\beta_I$.
The lines show the best-fitting $\beta_I$ and its 1$\sigma$ errors. }
\label{betaplot}
\end{figure}

A final caveat should be noted: the preliminary analysis presented here is based on a Method I 
approach and is therefore affected by Malmquist Bias. Homogeneous 
Malmquist effects have been 
corrected for and the effects of Inhomogeneous Malmquist Bias (IMB) are suppressed by the use
of a cluster sample with relatively small distance errors per `object'. Nonetheless, it may be
expected that the `spurious infall' patterns produced by IMB act to artificially inflate
$\beta_I$. A crude attempt to judge the effect can be made by removing the 12 clusters with
$|v_{PSCz}| > 500$\,km\,s$^{-1}$, which lie mostly in infall regions where IMB will be most severe. 
The result has much greater uncertainties of course, since we have removed the clusters which contribute
the greatest signal-to-noise: we obtain $\beta_I=0.6\pm0.4$, with no measurable residual bulk-flow 
(amplitude $100\pm200$\,km\,s$^{-1}$). In a future analysis of the SMAC--PSCz comparison, we
will employ a Method II approach which is free from Malmquist bias effects.

\section{The Great Attractor and Shapley Concentration}

The SMAC sample includes 10 clusters within 15$^\circ$ of the Shapley / Great Attractor (GA)
direction (visible in Figure~1 as the concentration of points at $l=315^\circ,b=30^\circ$). 
Since this region is of some historical interest, we compare in Figure~3 the 
observed velocities with some (over-)simplistic toy-models for the dynamics of the region. The
models are based upon the simple spherical attractors of Faber \& Burstein (1988). We consider 
the effect of two such structures, one at the position of the `traditional' GA (at a distance of 
4500\,km\,s$^{-1}$) and a second centered on the Shapley core region (at 14500\,km\,s$^{-1}$).
The models are normalised to generate the GA-directed component of the Local Group's velocity 
in the CMB frame, viz. 500\,km\,s$^{-1}$. 

The leftmost panel of Figure~3 demonstrates that a pure GA infall model is a poor
fit to the SMAC data, especially in the immediate GA background, where no `far-side infall' is 
observed. Equally, the extreme Shapley infall model dramatically over-predicts velocities beyond 
$\sim$10000\,km\,s$^{-1}$ (middle panel). Allowing for contributions from both attractors, and
fitting for their relative amplitudes as the single free parameter, we can obtain an acceptable 
fit to the data, as shown in the rightmost panel. The best-fitting model has Shapley and the GA each
generating $50\pm10$\% of the Local Group velocity in this direction, proportions which accord with 
Hudson's (1994) conclusions based on the Mark II dataset.

Here, as in the case of `bulk-flow' we caution against drawing quantitative conclusions from
analyses based on unrealistically simple models of the velocity field. However, the above fits draw 
attention to the qualitative behaviour of the SMAC velocity data in the GA / Shapley 
direction. The absence of far-side infall in the SMAC data (here visualised as a retardation
of one infall pattern by another more distant infall structure) is also apparent in the PSCz maps, 
which reveal a wealth of structure along the GA-Shapley axis. 

\section{Conclusions}

From our analyses of the SMAC velocity field, as presented in the preceeding sections, 
we draw the following conclusions: 
\begin{enumerate}
\item{The best-fitting bulk-flow model has amplitude $640\pm200$\,km\,s$^{-1}$,
towards $(l,b) = (260^\circ, -2^\circ)$.}
\item{The direction of the observed bulk-flow is within $\sim$15$^\circ$ of the
direction predicted from the IRAS--PSCz redshift survey for the SMAC sample, while the
amplitude can be matched if $\beta_I\sim1$.}
\item{Comparing observed and predicted velocities cluster-by-cluster yields 
$\beta=0.95\pm0.25$, compatible with other analyses based on the PSCz.}
\item{The observed velocity field is fully accounted for by the local 
density field, with no recourse to residual flow generated beyond the limits
of PSCz, or to `spurious' flows associated with systematic errors.}
\item{The Shapley Concentration may be responsible for a significant fraction of 
the infall traditionally associated with the Great Attractor.}

\end{enumerate}

\begin{figure}\vskip -10mm
\plotone{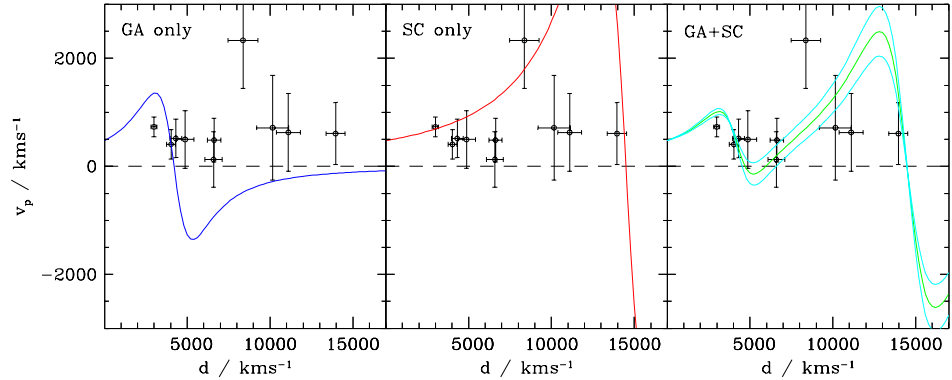}
\vskip -120mm
\caption{SMAC peculiar velocities in the Great Attractor / Shapley direction. 
The data are identical in each panel. The models shown are: a spherically symmetric 
Great Attractor model (left); a similar model attractor at Shapley (middle); the best 
fitting `two-attractor' model, with errors (right). All models are normalized to generate 
500\,km\,s$^{-1}$ at the Local Group.}\label{gashap}
\end{figure}

\acknowledgments

We thank Enzo Branchini, Carlos Frenk and the PSCz team 
for allowing us to discuss collaborative work in advance
of publication.


\begin{references}
\reference Burstein D., Heiles C. 1982, \aj, 87, 1165
\reference Dale D. A., Giovanelli R., Haynes M. P., Campusano L. E., Hardy E., 
	Borgani S., \apj, 510, L11
\reference Faber S. M., Burstein D., 1988, in Coyne G., Rubin V. C., eds,
    Proceedings of the Vatican Study Week, Large Scale Motions in the
    Universe. Princeton Univ. Press, Princeton, p.~135
\reference Feldman H. A., Watkins R. 1994, \apj, 430, L17
\reference Hudson M.~J. 1994, MNRAS, 266, 468
\reference Hudson M. J., Lucey J. R., Smith R. J., Steel J. 1997, \mnras, 291, 488
\reference Hudson M. J., Smith R. J., Lucey J. R., Schlegel D. J., Davies R. L. 1999, \apj, 512, L79
\reference Lauer T. R., Postman M. 1994, \apj, 425, 418 
\reference Schlegel D. J., Finkbeiner D. P., Davis M. 1998, \apj, 500, 525 
\reference Smith R. J., Lucey J. R. Hudson M. J., Steel J. 1997, \mnras, 291, 461
\reference Willick J. A. 1999, \apj, submitted (astro-ph/9812470)
\end{references}
\end{document}